\documentclass[12pt]{article}
\usepackage[dvips]{graphics}
\usepackage{epsfig}
\usepackage{latexsym}

\catcode`@=11
\def\@citex[#1]#2{\if@filesw\immediate\write\@auxout{\string\citation{#2}}\fi
  \def\@citea{}\@cite{\@for\@citeb:=#2\do
    {\@citea\def\@citea{,\penalty\@m}\@ifundefined
      {b@\@citeb}{{\bf ?}\@warning
       {Citation `\@citeb' on page \thepage \space undefined}}%
\hbox{\csname b@\@citeb\endcsname}}}{#1}}

\def\citer{\@ifnextchar
[{\@tempswatrue\@citexr}{\@tempswafalse\@citexr[]}}
\def\@citexr[#1]#2{\if@filesw\immediate\write\@auxout{\string\citation{#2}}\fi
  \def\@citea{}\@cite{\@for\@citeb:=#2\do
    {\@citea\def\@citea{--\penalty\@m}\@ifundefined
       {b@\@citeb}{{\bf ?}\@warning
       {Citation `\@citeb' on page \thepage \space undefined}}%
\hbox{\csname b@\@citeb\endcsname}}}{#1}}
\catcode`@=12

\def\bra#1{{\langle#1\vert}}
\def\ket#1{{\vert#1\rangle}}
\newcommand{\lsim}{\buildrel < \over {_\sim}}

\newcommand{\be}{\begin{equation}}
\newcommand{\ee}{\end{equation}}
\newcommand{\bea}{\begin{eqnarray}}
\newcommand{\eea}{\end{eqnarray}}
\newcommand{\ba}{\begin{array}}
\newcommand{\ea}{\end{array}}

\newcommand{\plb}[2]{{\em Phys. Lett.}              {\bf #1B}, #2 }
\newcommand{\npb}[2]{{\em Nucl. Phys.}              {\bf B#1}, #2 }
\newcommand{\npp}[2]{{\em Nucl. Phys. Proc. Suppl.} {\bf  #1}, #2 }
\newcommand{\pr }[2]{{\em Phys. Rep.}               {\bf  #1}, #2 }
\newcommand{\pra}[2]{{\em Phys. Rev.}               {\bf A#1}, #2 }
\newcommand{\prc}[2]{{\em Phys. Rev.}               {\bf C#1}, #2 }
\newcommand{\prd}[2]{{\em Phys. Rev.}               {\bf D#1}, #2 }
\newcommand{\prl}[2]{{\em Phys. Rev. Lett.}         {\bf  #1}, #2 }
\newcommand{\zpc}[2]{{\em Z. Phys.}                 {\bf C#1}, #2 }
\newcommand{\sci}[2]{{\em Science}                  {\bf  #1}, #2 }

\newcommand{\jhep}[2]{{\em J. High Energy Phys.}    {\bf  #1}, #2 }
\newcommand{\jpf}[2]{{\em J. Phys.} (France)        {\bf  #1}, #2 }

\newcommand{\rmp}[2]{{\em Rev. Mod. Phys.}          {\bf  #1}, #2 }

\newcommand{\nca}[2]{{\em Nuovo Cim.}               {\bf A#1}, #2 }
\newcommand{\con}[2]{                               {\bf  #1}, #2 }
\newcommand{\etal}{{\em et al.}}
\newcommand{\ibid}{{\em ibid.}}

\def\sinhat{\hat{s}^2}
\def\coshat{\hat{c}^2}
\def\sinzero{\sin^2\hat\theta_W (0)}
\def\alphat{\hat\alpha}
\def\qwp{Q_W(p)}
\def\qwn{Q_W(n)}
\def\qwe{Q_W(e)}
\def\qwcs{Q_W({\rm Cs})}
\def\qwtl{Q_W({\rm Tl})}
\def\sstw{\sin^2\theta_W}

\def\bra#1{\langle#1\vert}
\def\ket#1{\vert#1\rangle}

\begin{document}

\hfill
\begin{tabular}{r}
{\normalsize FT--2003--02} \\
{\normalsize Caltech MAP--287}
\end{tabular}

\vspace{18pt}
\centerline{\Large\bf The Weak Charge of the Proton and New Physics}
\vspace{18pt}
\centerline{\sc Jens Erler$^{\,a,b}$, Andriy Kurylov$^{\,c}$, and 
                Michael J. Ramsey-Musolf$^{\,b,c,d}$}
\vspace{6pt}
\centerline{\it $^a$Instituto de F\'\i sica, Universidad Nacional Aut\'onoma
                de M\'exico, 04510 M\'exico D.F., M\'exico}
\centerline{\it $^b$Institute for Nuclear Theory, University of Washington, 
                Seattle, WA 98195, USA}
\centerline{\it $^c$Kellogg Radiation Laboratory, California Institute of
                Technology, Pasadena, CA 91125, USA}
\centerline{\it $^d$Department of Physics,
                University of Connecticut, Storrs, CT 06269, USA}
\vspace{6pt}

\centerline{{\sl e-mail:} {\rm erler@fisica.unam.mx, kurilov@krl.caltech.edu,
mjrm@krl.caltech.edu}}
\vspace{18pt}

\begin{abstract}
We address the physics implications of a precision determination of the weak 
charge of the proton, $\qwp$, from a parity violating elastic electron proton
scattering experiment to be performed at the Jefferson Laboratory.  We present 
the Standard Model (SM) expression for $\qwp$ including one-loop radiative 
corrections, and discuss in detail the theoretical uncertainties and missing 
higher order QCD corrections. Owing to a fortuitous cancellation, the value of 
$\qwp$ is suppressed in the SM, making it a unique place to look for physics 
beyond the SM. Examples include extra neutral gauge bosons, supersymmetry, and
leptoquarks. We argue that a $\qwp$ measurement will provide an important 
complement to both high energy collider experiments and other low energy 
electroweak measurements. The anticipated experimental precision requires 
the knowledge of the ${\cal O}(\alpha_s)$ corrections to the pure electroweak 
box contributions. We compute these contributions for $\qwp$, as well as for 
the weak charges of heavy elements as determined from atomic parity violation. 
\end{abstract}

\vspace{0pt}

\section{Introduction}
\label{intro}
Precision tests continue to play an essential role in elucidating the structure
of the electroweak (EW) interaction~\citer{Langacker:1995qb,Erler:2002}. Such 
tests include the completed high energy program on top of the $Z$ resonance at 
the $e^+e^-$ accelerators, LEP~1 and SLC; precision measurements at LEP~2 and 
the $p\bar{p}$ collider Tevatron; and deep inelastic scattering (DIS) at 
the $ep$ collider HERA. Recent precision measurements at lower energies, such 
as a determination of the muon anomalous magnetic moment, 
$a_\mu$~\cite{Bennett:2002jb}, and of cross sections for neutrino-nucleus 
DIS~\cite{Zeller:2001hh}, have shown deviations from the SM expectations and 
generated some excitement about possible signatures of new physics, although 
theoretical uncertainties from the strong interaction presently cloud 
the interpretation of the results~\citer{Knecht:2001qf,Davidson:2001ji}.

In this paper we focus on the prospective impact of a precision low energy
measurement of the weak charge of the proton, $\qwp$, using parity violating 
(PV) elastic $ep$ scattering.  Such an experiment has recently been
proposed~\cite{Armstrong:2001} and approved at the Thomas Jefferson National 
Accelerator Facility (JLab) using the Continuous Electron Beam Accelerator 
Facility (CEBAF). Historically, semileptonic neutral current experiments have 
contributed substantially to our understanding of the EW interaction. In 
particular, the deep inelastic $eD$ asymmetry measurement at 
SLAC~\cite{Prescott:1979dh} in the late 1970's played a crucial role in 
singling out the SM over its alternatives at that time, and provided first 
measurements of the effective PV electron-quark couplings, 
$2\, C_{1u} - C_{1d}$, and, $2\, C_{2u} - C_{2d}$ (defined in 
Section~\ref{4fermi}). Subsequently, the latter combination was determined more
precisely in DIS of muons from carbon at CERN~\cite{Argento:1982tq}. 
Quasielastic and elastic electron scattering, respectively, from $^9$Be at 
Mainz~\cite{Heil:dz} and $^{12}$C at MIT-Bates~\cite{Souder:ia}, constrained 
the remaining linear combinations. More recently, measurements of the elastic 
$ep$ and $eD$ asymmetries at MIT-Bates~\cite{Hasty:2001ep} and 
JLab~\cite{Aniol:2000at} have been used to derive information on the neutral 
weak magnetic, electric, and axial vector form factors of the proton at 
$q^2 \neq 0$, and yielded a value for $C_{2u} - C_{2d}$~\cite{Hasty:2001ep}. 
Experiments probing atomic parity violation (APV) provided further precise 
information on various linear combinations of 
the $C_{1i}$~\citer{Edwards:1995,Wood:1997zq}. On the other hand, the neutral 
weak charge of the proton, proportional to $2\, C_{1u} + C_{1d}$, has never 
been measured.

In its own right, $\qwp$ is a fundamental property of the proton, being the
neutral current analog of the vector coupling, $G_V$, which enters neutron and 
nuclear $\beta$ decay. While measurements of $G_V$ provide the most precise 
determination of the CKM matrix element, $V_{ud}$, a precise determination of 
$\qwp$ may provide insight into the SM and its possible extensions. Because 
the value of the weak mixing angle, $\sin^2\theta_W$, is numerically close to 
1/4,
\be
   \qwp = 1 - 4\sin^2\theta_W,
\label{QW}
\ee
is suppressed in the SM (see Section~\ref{SM}). This suppression is 
characteristic for protons (and electrons) but not neutrons, and therefore it 
is absent in any other nucleus. As a consequence, $\qwp$ is unusually sensitive
to $\sin^2\theta_W$ and offers a unique place to extract it at low momentum 
transfer. Doing so will provide a test of the renormalization group evolution 
(RGE) of $\sin^2\theta_W$.

To put this statement in context, we note that the strong coupling, $\alpha_s$,
is routinely  subjected to analogous RGE tests, whose results provide crucial
evidence that QCD is the correct theory of the strong interaction. As we 
discuss in Section~\ref{SM}, a precise measurement of $\qwp$ --- along with 
the analogous measurement of the weak charge of the electron, $\qwe$, currently
measured by the E-158 Collaboration at SLAC~\cite{Carr:1997fu} --- will provide
this important test for the EW sector of the theory. An observed deviation of
the running of $\sin^2\theta_W$ from the SM prediction could signal 
the presence of new physics, whereas agreement would place new constraints on 
possible SM extensions. This test has taken on added interest recently in light
of the $\nu$-nucleus DIS results obtained by the NuTeV 
Collaboration~\cite{Zeller:2001hh} which show a 3~$\sigma$ deviation from 
the SM prediction. In contrast, the most recent determination of the weak 
charge of cesium, $\qwcs$, obtained in an APV experiment at 
Boulder~\cite{Wood:1997zq}, agrees with the SM value for this quantity and 
confirms the predicted SM running. However, the interpretation of both 
the cesium and NuTeV results has been subject to debate. For example, 
the extraction of $\qwcs$ from the experimental PV amplitude relies on 
intricate atomic structure 
computations~\citer{Derevianko:2000dt,Milstein:2002ai}, and the level of 
agreement with the SM has varied significantly as additional atomic structure 
effects have been incorporated in the calculations (see Section~\ref{general} 
for a discussion). Similarly, the NuTeV discrepancy may result from previously
unaccounted for effects in parton distribution 
functions~\cite{Bodek:1999bb,Davidson:2001ji}. At present, there are no other 
determinations of $\sin^2\theta_W$ off the $Z$ peak which have comparable 
precision.

Our discussion of the physics of $\qwp$ is organized as follows. 
In Section~\ref{general}, we review some general considerations of the PV $ep$ 
asymmetry and how $\qwp$ is extracted from it. We argue that this will be 
a theoretically cleaner procedure than the current extraction of $\qwcs$ from 
APV. Section~\ref{SM} gives details of the SM prediction for $\qwp$, which 
provides the baseline for comparison with experiment. Section~\ref{4fermi} is 
devoted to the prospective model independent constraints the new $\qwp$ 
experiment would generate. In Sections~\ref{zprimes} and~\ref{susy} we analyze
the sensitivity of $\qwp$ to extra neutral gauge bosons, supersymmetry (SUSY),
and leptoquarks (LQs). We summarize our conclusions in Section~\ref{conclude}.

\section{Parity violating $ep$ scattering and $\qwp$}
\label{general}
The PV $ep$ asymmetry has the simple form~\cite{Musolf:1993tb},
\be
\label{eq:alr}
A_{LR} = {{\sigma_{L} - \sigma_{R}} \over {\sigma_{L} + \sigma_{R}}}
       = -{G_F Q^2 \over 4\sqrt{2}\pi\alpha} \left[ \qwp + F^p(Q^2,\theta) 
         \right],
\ee
where $G_F$ is the Fermi constant, $Q^2$ is the momentum transfer, and $F^p$ is
a form factor. At forward angles, one has $F^p=Q^2 B(Q^2)$, where $B(Q^2)$ 
depends on the nucleon, electromagnetic (EM), and strangeness form factors. 
The present program of PV $ep$ scattering experiments --- which involve 
measurements~\cite{Hasty:2001ep,Aniol:2000at,Maas:1997ux,Arvieux:2001nr} of 
$A_{LR}$ over a wide kinematic range --- is designed to determine $F^p$ for 
forward angles at $Q^2$ values as low as $\sim 0.1 \mbox{ GeV}^2$. 
The determination of $\qwp$ involves an additional $A_{LR}$ measurement at 
$Q^2 \sim 0.03 \mbox{ GeV}^2$. Such a value of $Q^2$ is optimal for separating 
$\qwp$ from $F^p$ with sufficient precision, while retaining sufficient 
statistics (note that $A_{LR}$ is itself proportional to $Q^2$). The E-158 
experiment is being carried out at almost the same value of $Q^2$.

An important feature of the asymmetry in Eq.~(\ref{eq:alr}) is its 
interpretability. Current conservation implies that $\qwp$ is protected from 
large strong interaction corrections involving the low energy structure of 
the proton. As we note in Section~\ref{SM}, residual strong interaction 
corrections involving, {\em e.g.}, two boson exchange box diagrams, are 
suppressed at $Q^2 = 0$. Effects that depend on $Q^2$ are included in $F^p$ and
will be constrained by the aforementioned program of experiments, thereby 
eliminating the need for a first principles nucleon structure calculation. 
Based on present and future measurements, the extrapolation of $F^p$ to 
$Q^2 = 0$ is expected to induce a 2\% uncertainty and will thus be considered 
part of the experimental error budget\footnote{In practice, this extrapolation 
can be implemented using chiral perturbation theory. Present and future 
measurements will determine all the relevant low energy constants.}.

In this respect, the extraction of $\qwp$ from $A_{LR}$ is complementary to 
the recent determination of $\qwcs$ in APV. The latter relies on an advanced 
atomic theory calculation of the small PV $6s \rightarrow 7s$ transition 
amplitude. Experimentally, the transition amplitude has been 
measured~\cite{Wood:1997zq} to a relative precision of 0.35\%. Subsequently, by
measuring the ratio of the off-diagonal hyperfine amplitude (which is known
precisely~\cite{Bouchiat:1988}) to the tensor transition
polarizability~\cite{Bennett:1999pd}, it was possible to determine $\qwcs$ with
a combined experimental and theoretical uncertainty of 0.6\%. The result 
differed by 2.3~$\sigma$ from the SM prediction~\cite{Erler:2002} for $\qwcs$. 
However, updating the corrections from the Breit 
interaction~\citer{Derevianko:2000dt,Kozlov:2001bn} and to a lesser degree from
the neutron distribution~\cite{Derevianko:2000dt,Pollock:1999ec} reduced 
the difference to only 1.0~$\sigma$, seemingly removing the discrepancy. 
Subsequent calculations included other large and previously underestimated 
contributions ({\em e.g.}, from QED radiative corrections), some 
increasing~\citer{Milstein:2001hi,Dzuba:2002hr}, others 
decreasing~\cite{Kuchiev:2002fg,Milstein:2002ai} the deviation. The atomic 
theory community now appears to agree on a 0.5\% atomic structure uncertainty 
for $\qwcs$, and in what follows we adopt the value,
\be
   Q_W({\rm Cs}) = -72.69 \pm 0.48.
\label{eq:qwcs}
\ee
There is also a noteworthy but less precise determination in 
Tl~\cite{Edwards:1995,Vetter:vf}, $\qwtl = -116.6 \pm 3.7$.

A possible strategy for circumventing atomic theory uncertainties is to measure
APV for different atoms along an isotope chain. Isotope ratios, ${\cal R}$, are
relatively insensitive to details of atomic structure and the attendant 
theoretical uncertainties, making them attractive alternatives to the weak 
charge of a single isotope as a new physics probe. As shown in 
Ref.~\cite{Ramsey-Musolf:1999qk}, any shift in ${\cal R}$ from its SM value due
to new physics would be dominated by the change in $\qwp$, as the effects on
${\cal R}$ of new physics corrections to the weak charge of the neutron,
$\qwn$, are suppressed. Moreover, ${\cal R}$ receives important contributions
from changes in the neutron distribution along the isotope
chain~\citer{Pollock:1999ec,Panda:pd}. At present, the corresponding nuclear 
structure uncertainties seem larger than needed to make ${\cal R}$ a useful 
probe of new physics effects on $\qwp$. In contrast, $ep$ scattering will yield
$\qwp$ without nuclear structure complications.

Given the suppression of $\qwp$ in the SM tree level expression~(\ref{QW}),
a 4\% measurement would provide a theoretically clean probe of new physics with
a sensitivity comparable to that achieved by a 0.5\% total error in $\qwcs$, 
but with entirely different systematical and theoretical uncertainties. Note, 
however, that measurements of $A_{LR}$ and single isotope APV are complementary
as they probe different combinations of the $C_{1i}$.  For example, in contrast
to the weak charges of heavy elements, $\qwp$ depends significantly on 
the oblique parameter $T$, introduced in Ref.~\cite{Peskin:1990zt}.

\section{$\qwp$ in the Standard Model}
\label{SM}
At tree level in the SM, $\qwp$ is given by Eq.~(\ref{QW}). Including radiative
corrections one can write,
\be
   \qwp = [ \rho_{NC} + \Delta_e ] [ 1 - 4 \sinzero + \Delta_e^\prime ]
        + \Box_{WW} + \Box_{ZZ} + \Box_{\gamma Z}.
\label{QWSM}
\ee
The parameter, $\rho_{NC} = 1 + \Delta\rho$~\cite{Veltman:1977kh}, renormalizes
the ratio of neutral to charged current interaction strengths at low energies,
and is evaluated including higher order 
QCD~\citer{Djouadi:1987gn,Chetyrkin:1995ix} and 
EW~\citer{Barbieri:1992nz,Degrassi:1996mg} corrections. We also include 
relatively small electron vertex and external leg corrections, which are 
corrections to the axial-vector $Zee$ and $\gamma ee$ couplings, 
respectively~\cite{Marciano:1983mm},
\be
   \Delta_e = - {\alpha\over 2\pi}, \hspace{100pt}
   \Delta_e^\prime = - {\alpha\over 3\pi} (1 - 4\sinhat) \left[
      \ln \left( {M_Z^2\over m_e^2} \right) + {1\over 6} \right].
\ee
The latter, which corresponds to the anapole moment of the electron, depends on
the choice of EW gauge and is not by itself a physical
observable~\cite{Musolf:1990sa}. The purely weak box contributions are given 
by~\cite{Marciano:1983mm,Musolf:ts},
\be
   \Box_{WW} = {7 \alphat \over 4 \pi \sinhat}, \hspace{100pt}
   \Box_{ZZ} = {\alphat\over 4\pi\sinhat\coshat} \left( {9\over 4} -
      5 \sinhat \right) (1 - 4\sinhat + 8\hat{s}^4),
\label{eq:wzbox}
\ee
where $\alphat \equiv \alphat (M_Z)$ and
$\sinhat \equiv 1 - \coshat \equiv \sin^2\hat\theta_W (M_Z)$ are 
the $\overline{\rm MS}$ renormalized QED coupling and weak mixing angle at 
the $Z$ scale, respectively. Numerically, the $WW$ box amplitude generates an 
important 26\% correction to $\qwp$, while the $ZZ$ box effect is about $3\%$.

These diagrams are dominated by intermediate states having 
$p^2 \sim {\cal O}(M_{W,Z}^2)$. The corresponding QCD corrections are, thus, 
perturbative and can be evaluated by relying on the operator product expansion
(OPE). At short distances, the product of weak currents entering the hadronic 
side of the box graphs is equivalent to a series of local operators whose 
Wilson coefficients can be evaluated by matching with a free field theory
calculation. Because the weak (axial) vector current is (partially) conserved,
the resulting operators have no anomalous dimensions. Consequently, 
the perturbative QCD (pQCD) contributions introduce no large logarithms.

In order to evaluate the ${\cal O}(\alpha_s)$ corrections to these graphs, we 
follow Ref.~\cite{Sirlin:1977sv} where analogous corrections for neutron 
$\beta$ decay are computed. For the $WW$ box graphs, we have the amplitude,
\begin{equation}
\label{eq:wwbox1}
  i {\cal M}_{WW} = i\left({g\over 2\sqrt{2}}\right)^4\int {d^4k\over (2\pi)^4}
  {\bar e}(K') \gamma^\nu (1-\gamma_5) \hspace{-5pt} \not k 
  \gamma^\mu(1-\gamma_5) e(K) T_{\mu\nu}(k) {1\over k^2}{1\over (k^2-M_W^2)^2},
\end{equation}
where,
\begin{equation}
  T_{\mu\nu}(k) = \int d^4x {\rm e}^{-ik\cdot x}\langle p'| T\left( J_\mu^+(0)
  J_\nu^-(x)\right)|p\rangle,
\end{equation}
with $J^\pm_\mu(x)$ being the charge changing weak currents. Since the loop
integral is infrared finite and is dominated by intermediate states having 
$k\sim M_W$, we have dropped all dependence on $m_e$ and the electron momenta 
$K$ and $K'$. The error introduced by this approximation is of order 
$(E_e/M_W)^2 \sim 0.02\% $ for the kinematics of the planned experiment, and is
negligible for our purposes. A little algebra allows us to rewrite 
Eq.~(\ref{eq:wwbox1}) as,
\begin{eqnarray}
\label{eq:wwbox2}
  i {\cal M}_{WW} & = & 2i\left({g\over 2\sqrt{2}}\right)^4 
  \int {d^4k\over (2\pi)^4} {\bar e}(K') 
  [k^\nu\gamma^\mu + k^\mu\gamma^\nu - g^{\mu\nu} \hspace{-5pt} \not k \\
\nonumber
  & & +i \epsilon^{\mu\nu\alpha\lambda}\gamma_\lambda\gamma_5 k_\alpha]
  (1-\gamma_5) e(K) T_{\mu\nu}(k) {1\over k^2} {1\over (k^2-M_W^2)^2}.
\end{eqnarray}
The terms proportional to $k^\mu T_{\mu\nu}$ and $k^\nu T_{\mu\nu}$ are
protected from large pQCD corrections by symmetry considerations. This feature 
may be seen by observing,
\begin{equation}
\begin{array}{c}
  k^\nu T_{\mu\nu} = \int d^4x (i\partial^\nu{\rm e}^{-ik\cdot x})
  \langle p'| T\left( J_\mu^+(0) J_\nu^-(x)\right)|p\rangle \vspace{8pt} \\
  = i\int d^4x {\rm e}^{-ik\cdot x}\delta(x_0) 
  \langle p'| [ J_\mu^+(0),J_0^-(x)] |p\rangle 
  -i \int d^4x {\rm e}^{-ik\cdot x} 
  \langle p'| T\left( J_\mu^+(0) \partial^\nu J_\nu^-(x)\right)|p \rangle,
\end{array}
\end{equation}
after integration by parts. The divergence $\partial^\nu J_\nu^-(x)$ vanishes 
in the chiral limit, and in keeping with the high-momentum dominance of 
the integral, may be safely neglected. On the other hand, the equal time 
commutator gives $-4i\bra{p'} J_\mu^3(0)\ket{p}$, where 
$J_\mu^3={\bar q}_L\gamma_\mu\tau_3 q_L$ and $q=(u,d)$. Note that 
the commutator term results from the $SU(2)_L \times U(1)_Y$ symmetry of
the theory, so it is not affected by QCD corrections.

In contrast, terms involving $T^\mu_\mu$ and 
$\epsilon^{\mu\nu\alpha\lambda} T_{\mu\nu}$ cannot be related to equal time 
commutators and, thus, involve {\em bona fide\/} short distance operator
products. In the OPE, the leading local operator appearing in $T^\mu_\mu$ is 
just $J_\mu^3$, whereas for the antisymmetric part, one has the isoscalar 
current, $J_\mu^{I=0} = {\bar q}_L\gamma_\mu q_L$. The leading pQCD 
contributions to the corresponding Wilson coefficients have been worked out in 
Refs.~\citer{Sirlin:1977sv,Beg:vs}. For both, $T^\mu_\mu$ and
$\epsilon^{\mu\nu\alpha\lambda} T_{\mu\nu}$, the correction factor is
$1 - \alpha_s(k^2)/\pi$. Since the loop integrals are dominated by 
$k^2 \sim M_W^2$, one may approximate the impact on $i {\cal M}_{WW}$ by 
factoring $1 - \alpha_s(M_W^2)/\pi$ out of the corresponding parts of 
the integral in Eq.~(\ref{eq:wwbox2}). The error associated with this
approximation is of order $\alpha_s^2$ and is devoid of any large logarithms. 
The resulting expression for the $WW$ box contribution to $\qwp$ is,
\begin{equation}
\label{eq:wwbox3}
  \Box_{WW} = {{\hat\alpha}\over 4\pi\sinhat} 
  \left[ 2 + 5 \left( 1 - {\alpha_s(M_W^2)\over\pi} \right) \right],
\end{equation}
where the first term inside the square brackets arises from the equal time
commutator. Numerically, the ${\cal O}(\alpha_s)$ term yields an $\approx -3\%$
correction to $\Box_{WW}$, for an $\approx -0.7\%$ correction to $\qwp$. Higher
order pQCD corrections should be an order of magnitude smaller, so the error in
$\qwp$ associated with truncating at ${\cal O}(\alpha_s)$ is well below 
the expected experimental uncertainty. 

The calculation of pQCD corrections to $\Box_{ZZ}$ follows along similar lines.
In this case, however, all equal time commutators vanish, so that the entire
integral carries a $1 - \alpha_s(M_Z^2)/\pi$ correction factor. The resulting 
shift in $\qwp$ is $-0.1\%$, and higher order pQCD effects are negligible. For 
both $\Box_{WW}$ and $\Box_{ZZ}$ contributions from lower loop momenta 
($k^2 \ll M_W^2$) are associated with non-perturbative QCD effects. Such
contributions, however, carry explicit $(p/M_{W,Z})^2$ suppression factors,
where $p$ is an external momentum or mass. Taking, $p\sim E_e \sim 1$~GeV 
implies that these non-perturbative contributions are suppressed by at least 
a few $\times 10^{-4}$, so we may safely neglect them here. A similar 
conclusion applies to matrix elements of higher order operators in the OPE
analysis of $T_{\mu\nu}$ given above.

As a corollary, we have also computed the analogous correction to $\qwn$. 
Again, the $ZZ$ box contribution receives an overall factor,
$1 - \alpha_s(M_Z^2)/\pi$, while for the $WW$ box we obtain,
\begin{equation}
\label{eq:wwbox3n}
  \Box_{WW}^{(n)} = {{\hat\alpha}\over 4\pi\sinhat} 
  \left[ - 2 + 4 \left( 1 - {\alpha_s(M_W^2)\over\pi} \right) \right].
\end{equation}
Notice that the sum of Eqs.~(\ref{eq:wwbox3}) and~(\ref{eq:wwbox3n}) is also
corrected by an overall factor, $1 - \alpha_s(M_W^2)/\pi$, as is expected from
an isoscalar combination where no equal time commutator should be involved. 
The resulting shifts in the SM predictions for $\qwcs$ and $\qwtl$ are $-0.07$
and $-0.11$, respectively, or +0.1\%. 

In contrast, the $\gamma Z$ box contribution,
\be
   \Box_{\gamma Z} = {5\alphat\over 2 \pi} (1 - 4\sinhat) \left[
      \ln \left( {M_Z^2\over \Lambda^2} \right) + C_{\gamma Z}(\Lambda)
\right],
\label{gzbox}
\ee
contains some sensitivity to the low momentum regime. The scale, 
$\Lambda \sim {\cal O}(1\mbox{ GeV})$, appearing here denotes a hadronic 
cut-off associated with the transition between short and long distance 
contributions to the loop integral. The former are calculable and are dominated
by the large logarithm, $\ln M_Z^2/\Lambda^2$. At present, however, one cannot 
compute long distance contributions from first principles in QCD. Consequently,
we parameterize them by the constant $C_{\gamma Z}(\Lambda)$, whose 
$\Lambda$-dependence must cancel that associated with the short distance 
logarithm. We note that a similar situation arises in radiative corrections to
$G_V$ in neutron and nuclear $\beta$ decay, where the $\gamma W$ box diagram 
contains a short distance logarithm and a presently uncalculable long distance 
term, $C_{\gamma W}(\Lambda)$.

In the case of $\qwp$, the uncertainty associated with $C_{\gamma Z}(\Lambda)$
is suppressed by the $ (1 - 4\sinhat)$ prefactor\footnote{Additional 
contributions arise that are not suppressed by this factor, but are negligible 
as they go as $(E_e/M_Z)^2$.} in Eq.~(\ref{gzbox}). This factor arises from 
the sum of box and crossed-box diagrams, leading to an antisymmetric product of
the lepton EM and weak neutral
currents~\cite{Ramsey-Musolf:1999qk,Marciano:1983mm}. Since the resulting 
leptonic part of the box amplitude must be axial vector in character, only 
the vector part of the weak neutral current of the electron enters which is 
proportional to $1 - 4 \sinhat$. This result is quite general and independent 
of the hadronic part of the diagram. To estimate this uncertainty numerically,
we follow Ref.~\cite{Marciano:1993ep} setting $\Lambda = m_\rho$ and
$C_{\gamma Z}(m_\rho) = 3/2 \pm 1$, which translates into a $\pm 0.65\%$ 
uncertainty in $\qwp$. The central value for $C_{\gamma Z}(m_\rho)$ is from 
a free quark calculation. A more detailed analysis, taking into account 
contributions from intermediate excited states of the proton, is likely to 
shift $C_{\gamma Z}$, but we do not expect the change to be considerably larger
than the estimated  uncertainty. In any case, increasing the error bar on 
$C_{\gamma Z}$ by a factor of five would still imply an uncertainty in $\qwp$ 
below the expected experimental error. For comparison, we note that a change in
the value of $C_{\gamma W}(\Lambda)$ of similar magnitude would substantially 
affect the extraction of $|V_{ud}|^2$ from light quark $\beta$ decays, causing
the first row of the CKM matrix to deviate from unitarity by several standard 
deviations. Since the dynamics entering $C_{\gamma Z}$ and $C_{\gamma W}$ are 
similar, it appears unlikely that the uncertainty in $C_{\gamma Z}$ could 
differ significantly from $\pm 1$.

The remaining hadronic contribution to $\qwp$ arises from the low energy weak 
mixing angle, $\sinzero$, which is the EW analog of the EM coupling, $\alphat$.
The latter is measured very precisely in the Thomson limit ($q^2 = 0$), but 
hadronic contributions induce a sizable uncertainty for large $q^2$, and most 
importantly for $q^2 = M_Z^2$~\cite{Erler:2001hn}. Conversely, $\sinhat$ is 
measured precisely at the $Z$ pole, but hadronic loops induce an uncertainty 
for $q^2 = 0$, which is correlated but not identical to the one in $\alphat$. 
Note that effects due to $q^2 \neq 0$ are already taken into account 
experimentally via the $Q^2$ expansion and extrapolation of $F^p$ (see
Section~\ref{general}). One can then define,
$$
   \sinzero = \sinhat + \Delta\kappa_{\rm had}^{(5)} + {\alpha\over\pi}
              \left\{ {(1 - 4\sinhat)\over 12}
                      \left[ \sum_\ell \ln \left( {M_Z^2\over m_\ell^2} \right)
                             \left( 1 + {3\alpha\over 4\pi} \right)
                             + {135\alpha\over 32\pi} \right] \right.
$$
\be
             \left. - \left[ {7\coshat\over 4} + {1\over 24} \right]
                      \ln \left( {M_Z^2\over M_W^2} \right)
                    + {\sinhat\over 6} - {7\over 18} \right\},
\label{sinzero}
\ee
where the sum is over the charged leptons, and we find for the hadronic 
contribution,
\be
\label{eq:kappahad}
   \Delta\kappa_{\rm had}^{(5)} = (7.90 \pm 0.05 \pm 0.06) \times 10^{-3},
\ee
inducing a 0.4\% uncertainty in $Q_W(p)$. The first error in
Eq.~(\ref{eq:kappahad}) is correlated with the uncertainty in 
$\Delta\hat\alpha_{\rm had}^{(5)}(M_Z^2)$~\cite{Erler:1998sy}. The second error
is from the conversion to $\Delta\kappa_{\rm had}^{(5)}$ which induces
an uncertainty from the flavor separation of the $e^+ e^-$ annihilation and
$\tau$ decay data. This updates the value in Ref.~\cite{Marciano:1993ep},
$\Delta\kappa_{\rm had}^{(5)} = (7.96 \pm 0.58) \times 10^{-3}$. Note that
the uncertainty in $\Delta\kappa_{\rm had}^{(5)}$ is also related to the vacuum
polarization uncertainty~\cite{Davier:2002dy,Hagiwara:2002ma} in $a_\mu$. 
These correlations should be properly treated in global analyzes of precision 
data. With $\sinhat = 0.23112 \pm 0.00015$ from a SM fit to all current data, 
Eqs.~(\ref{QWSM}) and~(\ref{sinzero}) yield,
\be
   \sinzero = 0.23807 \pm 0.00017, \hspace{100pt} \qwp = 0.0716 \pm 0.0006,
\ee
where the uncertainty in the prediction for $\qwp$ is from the input parameters
and dominated by the error in $\sinhat$. The latter will decrease significantly
in the future~\cite{Baur:2002gp}. Taken together, the hadronic effects arising 
from $\Delta\kappa_{\rm had}^{(5)}$ and the box graphs combine to give 
a theoretical uncertainty of 0.8\%.

\begin{figure}
\centering
\rotatebox{0.}{\resizebox{5.7in}{7.0in}{\includegraphics{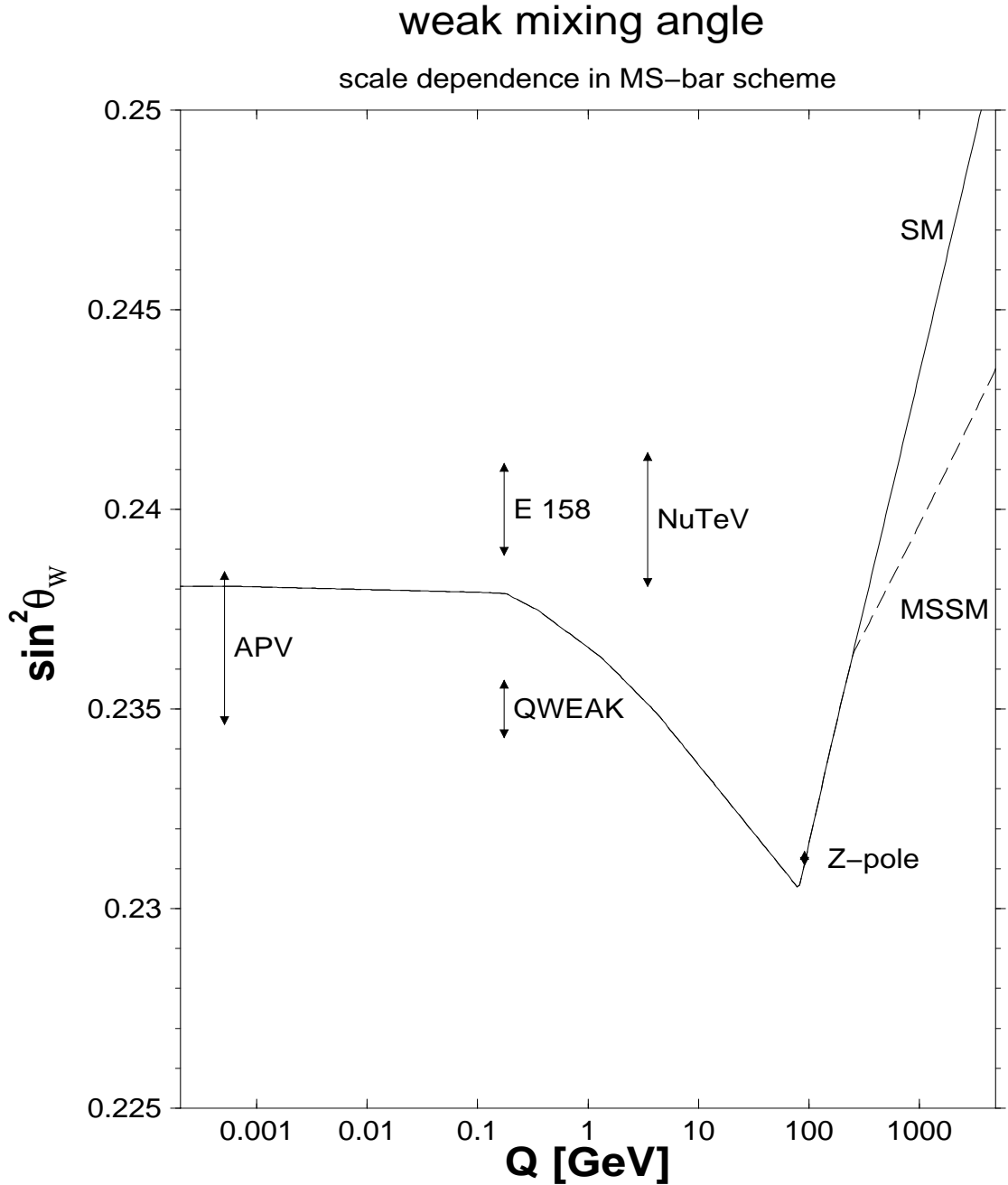}}}
\caption{Calculated running of the weak mixing angle in the SM, defined in
the $\overline{\rm MS}$ renormalization scheme (the dashed line indicates
the reduced slope typical for the Minimal Supersymmetric Standard Model). Shown
are the results from APV (Cs and Tl), NuTeV, and the $Z$-pole. QWEAK and E 158 
refer to the future $\qwp$ and $\qwe$ measurements and have arbitrarily chosen
vertical locations.}
\label{RUNNINGTHETA}
\end{figure}

The QWEAK experiment~\cite{Armstrong:2001} seeks to perform the most precise 
determination of the weak mixing angle off the $Z$ pole. For example, a 4\% 
determination, $\Delta \qwp = \pm 0.0029$~\cite{Armstrong:2001}, (assuming 
a 2.8\% statistical plus 2.8\% systematic plus 0.8\% theoretical error) would 
yield an uncertainty,
\be
   \Delta \sinzero = \pm 7.2 \times 10^{-4}.
\label{deltasin}
\ee
While the precise definition of $\sinzero$ is scheme dependent, this quantity 
is nonetheless useful for comparing different low energy experiments. 
Furthermore, as illustrated in Fig.~\ref{RUNNINGTHETA}, the $q^2$ evolution 
from the $Z$ pole as predicted by the SM,
\be
  \sinzero - \sinhat = 0.00694 \pm 0.00074,
\label{RGE_QWp}
\ee
could be established with more than 9 standard deviations. For comparison,
the cleanest test of pQCD can be obtained by contrasting the $\tau$ lepton 
lifetime with the hadronic $Z$ decay width: when interpreted as the RGE 
evolution of $\alpha_s$ from $m_\tau$ to $M_Z$, the result of the latest 
analysis~\cite{Erler:2002bu} corresponds to an 11~$\sigma$ effect.

Before proceeding, we comment on one additional possible source of hadronic
effects in $\qwp$: isospin admixtures in the proton wavefunction. The SM value 
quoted above implicitly assumes that the proton is an exact eigenstate of 
isospin. The EM and weak neutral vector currents for light quarks can then be 
decomposed according to their isospin content,
\begin{eqnarray}
  J_\mu^{EM} & = & \sum_{q=u,d} Q_q {\bar q}\gamma_\mu q 
               = J_\mu^{I=1}+J_\mu^{I=0}, \\
  J_\mu^{NC} & = & - 2 \sum_{q=u,d} C_{1q}{\bar q}\gamma_\mu q
               =  -2(C_{1u}-C_{1d}) J_\mu^{I=1} -6(C_{1u}+C_{1d}) J_\mu^{I=0},
\end{eqnarray}
where the $C_{1q}$ are defined in Eq.~(\ref{eq:lsmodel}). For the purpose of 
this discussion, we neglect contributions from strange quarks, which are 
effectively contained in $F^p$ term in Eq.~(\ref{eq:alr}). To the extent to 
which the nucleon is a pure $I = 1/2$ isospin eigenstate, one has 
$F_1^p(0)^{I=1} = F_1^p(0)^{I=0} =1/2$, where the $F_1^p(0)^I$ are the Dirac 
form factors associated with the proton matrix elements of the $J_\mu^I$.
In principle, these form factor relations receive small corrections due to
isospin breaking light quark mass differences ($m_u \neq m_d$) and EM effects.
However, conservation of EM charge implies that such corrections vanish. To see
this, assume that the proton state contains a small, ${\cal O}(\epsilon)$, 
admixture of an $I' \neq 1/2$ state,
\begin{equation}
  \ket{p} = \sqrt{1-\epsilon^2} \ket{1/2, 1/2} + \epsilon \ket{I', I_3'},
\end{equation}
where, for the purpose of this illustration, we drop explicit 
${\cal O}(\epsilon^2)$ terms involving the $\ket{I', I_3'}$ state. At $q^2=0$,
the charges, $J_0^I$, are equivalent to the operators, ${\hat I}_3$ and
$\frac{1}{2}{\hat 1}$. Since these operators cannot connect states of different
total isospin, one has,
\begin{eqnarray}
  F_1^p(0)^{I=1}& =& \frac{1}{2}(1-\epsilon^2)+ \epsilon^2I_3', \\
  F_1^p(0)^{I=0}& =& \frac{1}{2}.
\end{eqnarray}
Since the proton charge is $1 = F_1^p(0)^{I=1} + F_1^p(0)^{I=0}$, one must have
$I_3'=1/2$, so that there are no corrections to $F_1^p(0)^{I}$ through 
${\cal O}(\epsilon^2)$. Thus, one has to this order for the neutral current 
Dirac form factor,
\begin{equation}
  \qwp \equiv F_1^p(0)^{NC} = -2 \left( 2\, C_{1u} + C_{1d} \right),
\end{equation}
which is the same result obtained in the absence of any isospin impurities.
Similar arguments prevent the appearance of any higher order terms in
$\epsilon$.

\section{Four-Fermi operators and model independent analysis}
\label{4fermi}
Before considering the consequences for particular models of new physics, it is
instructive to consider the model independent implications of a 4\% $\qwp$ 
measurement. The low energy effective electron-quark Lagrangian of the form 
$A(e) \times V(q)$ is given by,
\begin{equation}
   {\cal L}={\cal L}^{\rm PV}_{\rm SM} + {\cal L}^{\rm PV}_{\rm NEW},
\end{equation}
where,
\begin{eqnarray}
\label{eq:lsmodel}
   {\cal L}^{\rm PV}_{\rm SM} &=& - {G_F \over \sqrt{2}} {\bar e} \gamma_\mu 
   \gamma_5 e \sum_q C_{1q}\ {\bar q} \gamma^\mu q, \\
\label{eq:lnew}
   {\cal L}^{\rm PV}_{NEW} &=& {g^2\over 4\Lambda^2} {\bar e} \gamma_\mu 
   \gamma_5 e \sum_f h_V^q\  {\bar q} \gamma^\mu q,
\end{eqnarray}
and where $g$, $\Lambda$, and the $h_V^q$ are, respectively, the coupling
constant, the mass scale, and effective coefficients associated with the new 
physics. The latter are in general of order unity; the explicit factor of 4
arises from the projection operators on left and right (or vector and 
axial-vector) chiral fermions. In the same normalization, the SM coefficients 
take the values, $C_{1u}/2 = - 0.09429 \pm 0.00011$ and 
$C_{1d}/2 = + 0.17070 \pm 0.00007$, for up and down quarks, respectively, where
we included the QCD corrections obtained in Eqs.~(\ref{eq:wwbox3}) 
and~(\ref{eq:wwbox3n}), and where the uncertainties are from the SM inputs.  
We find,
\begin{equation}
   Q_{W}^{p} ({\rm SM}) = - 2 (2 C_{1u} + C_{1d}) = 0.0716 \pm 0.0006.
\end{equation}
A 4\% measurement of $\qwp$ would thus test new physics scales up to,
\begin{equation}
   {\Lambda\over g} \approx {1\over\sqrt{\sqrt{2} G_F |\Delta Q_{W}^{p}|}}
                    \approx 4.6~\hbox{TeV}.
\end{equation}
The sensitivity to non-perturbative theories (such as technicolor, models of 
composite fermions, or other strong coupling dynamics) with $g\sim 2\pi$ could 
even reach $\Lambda \approx 29~\hbox{TeV}$. As another example, for extra $Z'$ 
bosons from simple models based on Grand Unified Theories (GUT) one expects 
$g \sim 0.45$, so that one can study such bosons (with unit charges) up to 
masses $M_{Z^\prime} \approx 2.1~\hbox{TeV}$. $Z'$ bosons are predicted in very
many extensions of the SM ranging from the more classical GUT and technicolor
models to SUSY and string theories. We discuss the sensitivity of $\qwp$ to
$Z'$ bosons, as well as other scenarios, in the subsequent Sections.

\begin{figure}
\centering
\rotatebox{0.}
{\resizebox{7.in}{7.in}{\includegraphics{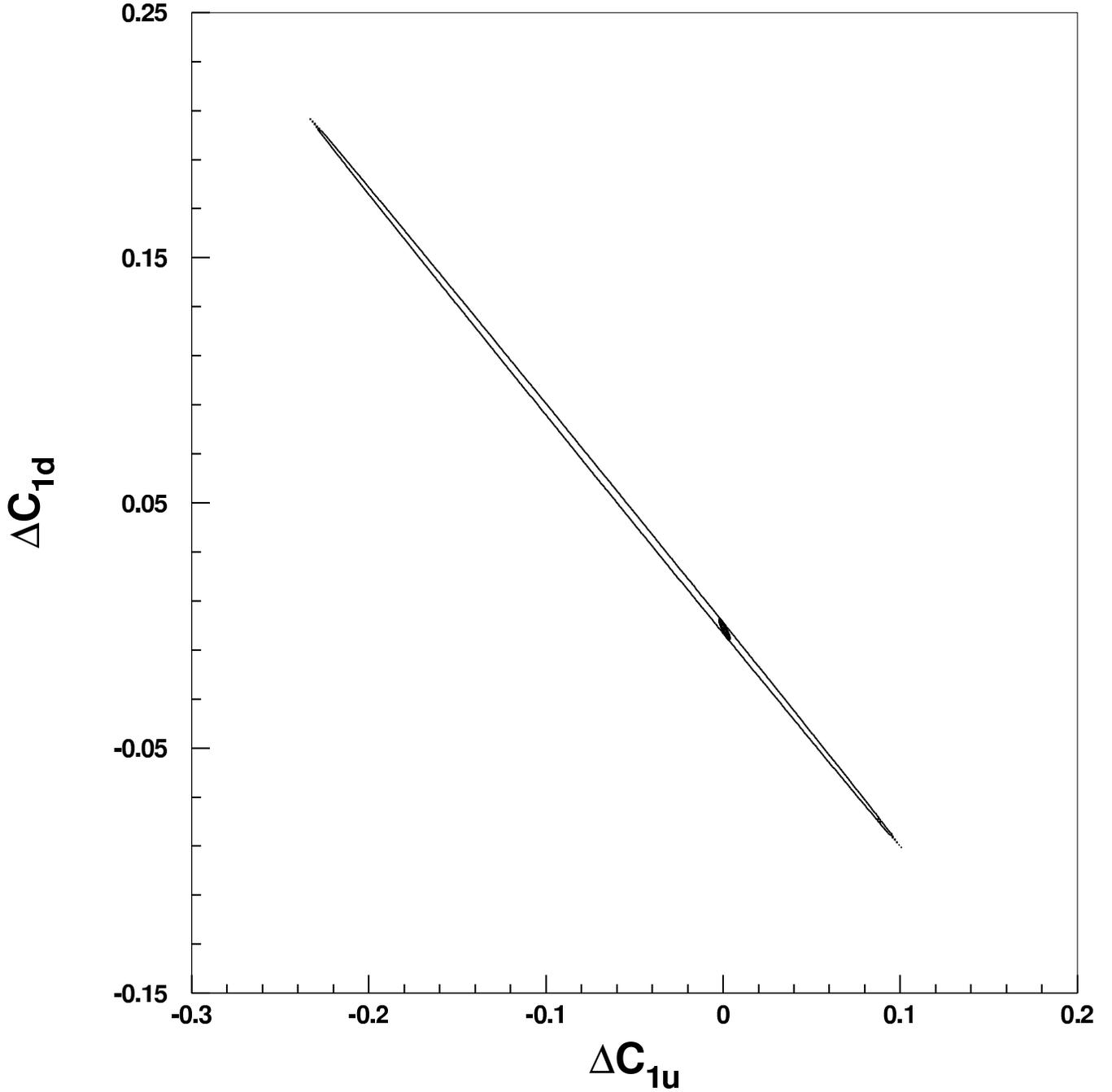}}}
\caption{Present and prospective 90\% C.L.\ constraints on new physics
contributions to the $eq$ couplings $C_{1u}$ and $C_{1d}$. The larger ellipse 
represents the present constraints, derived from APV in Cs~\cite{Wood:1997zq}, 
and polarized electron scattering at MIT-Bates~\cite{Souder:ia} and 
SLAC~\cite{Prescott:1979dh}. The smaller ellipse indicates the constraints 
after the inclusion of the $\qwp$ measurement, assuming the central 
experimental value coincides with the SM prediction. 
\label{c1dc1u}}
\end{figure}

In Fig.~\ref{c1dc1u} we plot the present constraints on $\Delta C_{1u}$ and
$\Delta C_{1d}$, the shifts in the $C_{1q}$ caused by new physics.  They are 
derived from $\qwcs$~\cite{Wood:1997zq}, as well as the MIT-Bates 
$^{12}$C~\cite{Souder:ia} and SLAC deuterium~\cite{Prescott:1979dh} parity 
violation measurements. As long as $\Delta C_{1u}$ and $\Delta C_{1d}$ are 
almost perfectly correlated, the result is an elongated ellipse. The impact of 
the proposed $\qwp$ measurement is indicated by the smaller ellipse. 
The dramatic reduction in the allowed parameter space will be possible because
$\qwp$ probes a very different linear combination than the existing data.

\begin{figure}
\begin{center}
\hspace{-55pt}
\rotatebox{0.}{\resizebox{7.5in}{7.5in}{\includegraphics{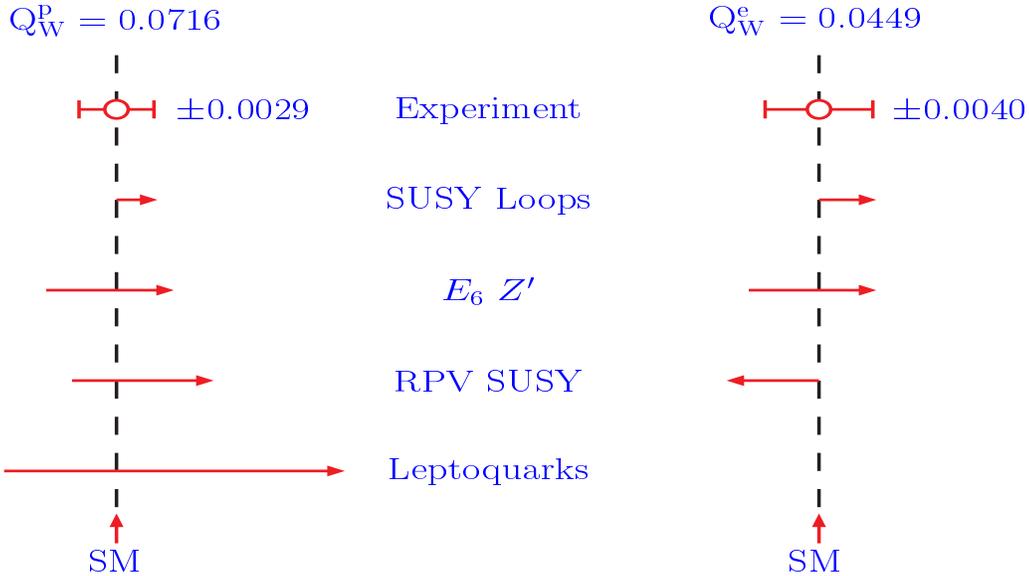}}}
\vspace{-280pt}
\end{center}
\caption{Comparison of anticipated errors for $\qwp$ and $\qwe$ with deviations
from the SM expected from various extensions and allowed (at 95\%~CL) by fits 
to existing data. Note that the two measurements are highly complementary. 
They would shift in a strongly correlated manner due to SUSY loops or a (1~TeV)
$Z'$ and thus together they could result in evidence for such new physics. 
In the case of RPV SUSY, the two measurements are somewhat anticorrelated. 
Finally, only $\qwp$ is sensitive to LQs, while $\qwe$ would serve as 
a control.}
\label{SMXsensitivity}
\end{figure}

In the next two Sections we turn to specific extensions of the SM, of which
there are many, and focus on three particularly well motivated types: gauge 
bosons, SUSY, and LQs.  In doing so, we emphasize the complementarity of the PV
M\o ller asymmetry measured by the SLAC-E-158 experiment~\cite{Carr:1997fu} 
which has comparable anticipated precision and (as a purely leptonic 
observable) has a clean theoretical interpretation. Some new physics scenarios
appear more strongly in the semileptonic channel than in the purely leptonic 
channel and {\em vice versa}. The complementarity of the two measurements is 
advantageous in attempting to distinguish among various new physics scenarios 
and is summarized in Fig.~\ref{SMXsensitivity}.

\section{Extra neutral gauge interactions}
\label{zprimes}
The introduction of neutral gauge symmetries beyond those associated with
the photon and the $Z$ boson have long been considered as one of the best
motivated extensions of the SM. Such $U(1)'$ symmetries are predicted in most 
GUTs and appear copiously in superstring theories. In the context of SUSY, they
do not spoil the approximate gauge coupling unification predicted by 
the simplest and most economic scenarios. Moreover, in many SUSY models
(though not the simplest $SO(10)$ ones), the enhanced $U(1)'$ gauge symmetry 
forbids an elementary bilinear Higgs $\mu$-term, while allowing an effective 
$\mu$ to be generated at the scale of $U(1)'$ breaking without introducing 
cosmological problems~\cite{Cvetic:1997wu}. In various string motivated models
of radiative breaking, this scale is comparable to the EW scale ({\em i.e.}, 
$\lsim 1$~TeV)~\cite{Cvetic:1997wu,Langacker:1998up}, thereby providing 
a solution~\cite{Cvetic:1995rj} to the $\mu$-problem~\cite{Kim:1984dt} and 
enhancing the prospects that a $Z'$ could be in reach in collider experiments 
or seen indirectly in the precision EW data. An extra $U(1)'$ symmetry could 
also explain proton stability, which is not automatic in supersymmetric models,
or it could solve both the proton lifetime puzzle and the $\mu$-problem
simultaneously~\cite{Erler:2000wu}.

From a phenomenological standpoint, direct searches at 
the Tevatron~\cite{Abe:1997fd} have as yet yielded no 
evidence\footnote{See, however, Ref.~\cite{Affolder:2001ha} which reports 
a 2~$\sigma$ deficit in the highest mass bin of the leptonic forward-backward 
asymmetry seen by the CDF Collaboration.} for the existence of an extra neutral
$Z'$ boson associated with the $U(1)'$, providing instead only lower bounds of
about 600~GeV (depending on the precise nature of the $Z'$). This implies 
a hierarchy of an order of magnitude between the $Z$ and $Z'$ masses. Recently,
using approximately flat directions in moduli space, it was shown that such 
a hierarchy can arise naturally in SUSY models~\cite{Erler:2002pr}.

On the other hand, several indirect effects could be attributed to a $Z'$.
The $Z$ lineshape fit at LEP~\cite{Abbaneo:2002} yields a significantly larger
value for the hadronic peak cross section, $\sigma^0_{\rm had}$, than is
predicted in the SM. This implies, for example, that the effective number of 
massless neutrinos, $N_\nu$, is $2.986 \pm 0.007$, which is 2~$\sigma$ lower 
than the SM prediction, $N_\nu = 3$. As a consequence, the $Z$ pole data
currently favors $Z'$ scenarios with a small amount of $Z$--$Z^\prime$-mixing 
($\sin\theta\neq 0$)~\cite{Erler:2000nx} which mimics a negative contribution 
to the invisible $Z$ decay width.  The result by the NuTeV 
Collaboration~\cite{Zeller:2001hh} can be brought into better agreement when 
one allows a $Z'$, especially when family non-universal couplings are 
assumed~\cite{Erler:2000nx,Langacker:2000ju}. 

To analyze the impact of a $Z'$ on $\qwp$, we employ Eq.~(\ref{eq:lnew}) with 
$\Lambda =M_{Z'}$ and $g = g_{Z^\prime} = 
\sqrt{5/3}\ \sin\theta_W \sqrt{\lambda}\ g_Z$~\cite{Langacker:1991pg}, where 
$\lambda = 1$ in the simplest models. $g_Z^2 = \sqrt{32}\ G_F M_Z^2$ is the SM
coupling constant for the ordinary $Z$. Consider the Abelian subgroups of 
the $E_6$ GUT group,
$$E_6 \to SO(10) \times U(1)_\psi \to SU(5) \times U(1)_\chi \times U(1)_\psi 
  \to SU(3)_C \times SU(2)_L \times U(1)_Y \times U(1)_\chi \times U(1)_\psi.$$
The most general $Z'$ boson from $E_6$ can be written as the linear
combination~\cite{Erler:2000nx},
\begin{equation}
  Z^\prime \sim - \cos\alpha \cos\beta\, Z_\chi + \sin\alpha \cos\beta\, Z_Y - 
  \sin\beta\, Z_\psi.
\label{eq:e6boson}
\end{equation}
Considerations of gauge anomaly cancellation as well as the proton lifetime and
$\mu$-problems in SUSY models mentioned earlier, also favor a $Z^\prime$ of 
that type~\cite{Erler:2000wu}. The assignment of SM fermions to representations
of $SO(10)$ implies that the $Z_\psi$ has only axial-vector couplings and can 
generate no PV $ef$ interactions of the type in Eq.~(\ref{eq:lnew}), whereas 
the $Z_\chi$ generates only PV $ed$ and $ee$ interactions of this type. 
Moreover, unlike in most other classes of models, the contributions to the weak
charges of the proton and the electron would have equal magnitude. Thus, should
$\qwp$ show a deviation from the SM prediction, a comparison with $\qwe$ would 
be a powerful tool to discriminate between a $Z'$ and other SM extensions. This
statement is illustrated in Fig.~\ref{SMXsensitivity} where the sensitivities 
of $\qwp$ and $\qwe$ are contrasted.

If a $Z'$ were detected at the Tevatron or the LHC, it would be important
to constrain its properties.  Its mass would be measured in course of 
the discovery, while $\sin\theta$ is mainly constrained by LEP~1. The $U(1)'$ 
charges and the couplings to quarks and leptons, however, are best determined 
by low energy precision measurements. Currently, the best fit values are, 
$\alpha = - 0.8^{+1.4}_{-1.2}$, $\beta = 1.0^{+0.4}_{-0.8}$, and
$\sin\theta = 0.0010^{+0.0012}_{-0.0006}$, obtained for $\lambda = 1$ and
$M_{Z'} = 1$~TeV. In this case, $\qwp = 0.0747$ is predicted, {\em i.e.\/} 
a 1.1~$\sigma$ effect. The impact of the QWEAK measurement would be to reduce
the allowed region of the parameters $\alpha$ and $\beta$ by $\sim 30\%$.

\section{Supersymmetry and leptoquarks}
\label{susy}
As in the case of extended gauge symmetry, the theoretical motivation for
supersymmetric extensions of the SM is strong. SUSY is a prediction of 
superstring theories; and if the SUSY breaking scale is at the EW scale, it 
stabilizes the latter and is consistent with coupling unification. Conversely,
minimal SUSY introduces a new set of issues, including the scale of the $\mu$ 
parameter mentioned above and the presence of 105 
parameters~\cite{Dimopoulos:1995ju,Haber:1997if} in the soft SUSY breaking 
Lagrangian. In order to be predictive, additional theoretical constraints must 
be invoked, such as those provided by gauge, gravity, or anomaly mediated SUSY 
breaking models. The phenomenological evidence for SUSY thus far is sparse, 
though hints exist. For example, the neutralino is a natural candidate for cold
dark matter, and the possible deviation of $a_\mu$ points suggestively toward 
SUSY. Since, in the end, experiment will determine what form of SUSY (if any) 
is applicable to EW phenomena, it is of interest to discuss the prospective 
implications of a $\qwp$ measurement for this scenario.

\begin{figure}
\begin{center}
\hspace*{-40pt}
\rotatebox{0.}{\resizebox{7.5in}{7.5in}{\includegraphics{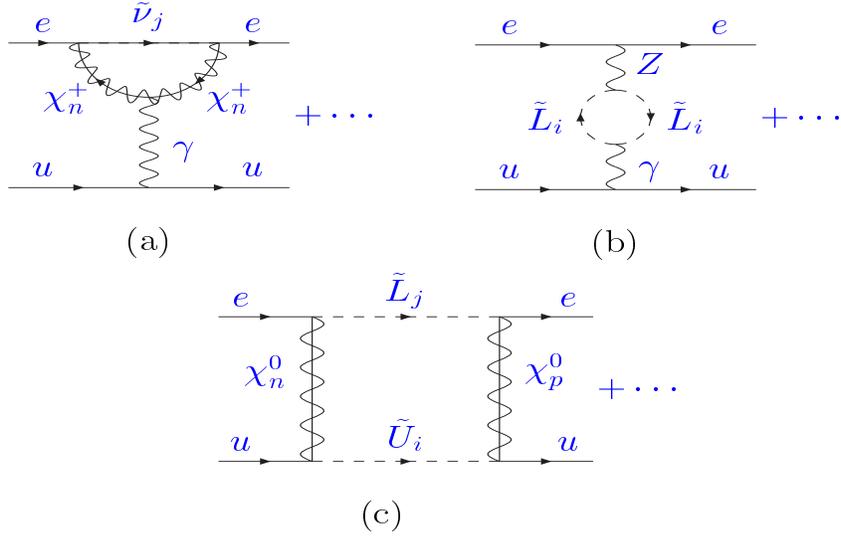}}}
\vspace{-340pt}
\end{center}
\caption{Representative examples of SUSY loop corrections to $\qwp$. Shown
are corrections from (a) charginos and sneutrinos; (b) sleptons contributing
to $\gamma$--$Z$-mixing ($\Delta \sin^2\hat\theta_W(0)_{\rm SUSY}$); and 
(c) a box graph containing neutralinos, sleptons, and squarks.}
\label{susyloops}
\end{figure}

While baryon number, $B$, and lepton number, $L$, are exact symmetries of 
the SM, they are not automatically conserved in the minimal supersymmetric 
Standard Model (MSSM). In order to avoid proton decay, $B$ and $L$ conservation
--- in the guise of $R$ parity conservation --- is often imposed by hand. 
In this case, every MSSM vertex contains an even number of superpartners, and 
the effects of SUSY appear in $\qwp$ only via loops, such as those shown in 
Fig.~\ref{susyloops}. Recently, such loop corrections to a variety of low and
medium energy precision observables were computed in 
Refs.~\citer{Kurylov:2001zx,Kurylov:2003by}. These analyzes were completed 
without invoking any assumptions about the mechanism for soft SUSY breaking. 
The implications of charged current data for the SUSY spectrum appear 
to conflict with those derived from typical models for SUSY breaking mediation.
This conflict may be alleviated by allowing for $R$ parity violation
(RPV)~\cite{Ramsey-Musolf:2000qn}, though doing so would eliminate the lightest
neutralino as a dark matter candidate. From this perspective, independent 
low energy probes of the MSSM spectrum take on added importance.

A measurement of $\qwp$, when considered in tandem with $\qwe$ and $\qwcs$, 
could provide such a probe. The MSSM loop corrections to the weak charges can 
be analyzed efficiently by modifying Eq.~(\ref{QWSM}),
\begin{equation}
\ba{c}
   \qwp = [\rho_{NC} + \Delta_e +\Delta\rho_{\rm SUSY}] [1 - 4 \sinzero + 
             \Delta_e^\prime ] + \Box_{WW} + \Box_{ZZ} + \Box_{\gamma Z} + 
             \lambda_{\rm SUSY}, \vspace{10pt} \\
   \sinzero = {\sin^2\hat\theta_W(0)}_{\rm SM} + 
                 \Delta \sin^2\hat\theta_W(0)_{\rm SUSY},
\ea
\end{equation}
where ${\sin^2\hat\theta_W(0)}_{\rm SM}$ is the SM prediction given in 
Eq.~(\ref{sinzero}) and $\Delta \sin^2\hat\theta_W(0)_{\rm SUSY}$ is 
the correction induced by SUSY loops\footnote{In the notation of 
Ref.~\cite{Kurylov:2002cz}, $\Delta \sin^2\hat\theta_W(0)_{\rm SUSY} = 
4 \hat{s}^2 \delta \kappa_{PV}^{\rm susy}$.}. All SUSY box graph contributions,
as well as non-universal vertex and external leg corrections, are contained in
$\lambda_{\rm SUSY}$. Flavor-independent corrections are given by 
$\Delta\rho_{\rm SUSY}$ and $\Delta \sin^2\hat\theta_W(0)_{\rm SUSY}$.

The effects of SUSY loops on $\qwp$ and $\qwe$ are dominated by 
$\Delta \sin^2\hat\theta_W(0)_{\rm SUSY}$, because present bounds on the $T$ 
parameter from precision data~\cite{Erler:2002} limit the magnitude of 
$\Delta\rho_{\rm SUSY}$. Moreover, box graph contributions are numerically 
small, while cancellations reduce the impact of vertex and external leg 
corrections. Consequently, the shifts in the proton and electron weak charges 
are similar over nearly all allowed SUSY parameter space. This is in contrast 
to $\qwcs$ due to canceling corrections to the $u$ and $d$ quark weak charges.
Thus, should the QWEAK and SLAC-E-158 experiments observe a correlated 
deviation, and should $\qwcs$ remain in agreement with the SM, the MSSM would 
be a favored explanation compared to many other scenarios.

\begin{figure}
\begin{center}
\hspace{-40pt}
\rotatebox{0.}{\resizebox{7.5in}{7.5in}{\includegraphics{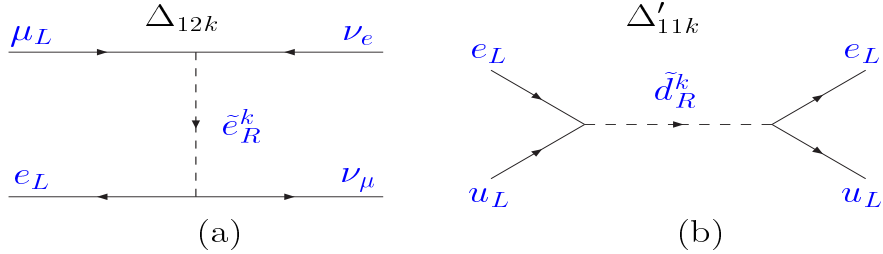}}}
\vspace{-410pt}
\end{center}
\caption{Representative examples of tree level SUSY corections in the case of
PRV. Shown are (a) a contribution to $\mu$ decay which affects $\qwp$ through
a modification of $G_F$, and (b) squark exchange.}
\label{rpvsusy}
\end{figure}

The situation changes considerably in the presence of RPV effects. The most 
general gauge invariant, renormalizable RPV extension of the MSSM is generated
by the superpotential~\cite{Hall:1983id},
\begin{equation}
\label{eq:rpvsp}
   W_{RPV} = \frac{1}{2} \lambda_{ijk} L^i L^j {\bar e}^k 
           + \lambda^\prime_{ijk} L^i Q^j {\bar d}^k 
           + \frac{1}{2} \lambda''_{ijk} {\bar u}^i {\bar d}^j {\bar d}^k
           + \mu^\prime_i L^i H_u,
\end{equation}
where $L^i$ and $Q^i$ denote the left-handed lepton and quark doublet 
superfields, respectively; the barred quantities denote the right-handed 
singlet superfields; $H_u$ is the hypercharge $Y = 1$ Higgs superfield; and 
the indices indicate generations. The bulk of studies of $W_{RPV}$ have been
phenomenological~\cite{Allanach:1999ic}. The strongest constraint comes from
the proton lifetime, which generally forbids the $B$ violating $\lambda''$ 
terms unless all other ($L$ violating) terms in $W_{RPV}$ vanish. Consequently,
we restrict our attention to $\lambda''_{ijk} = 0$, and for simplicity, we also
set $\mu'_i = 0$. When inserted into the amplitudes of Fig.~\ref{rpvsusy}, 
the remaining interactions in Eq.~(\ref{eq:rpvsp}) generate corrections in 
terms of the quantities $\Delta_{ijk}({\tilde f})$ and 
$\Delta_{ijk}'({\tilde f})$, where for example,
\begin{equation}
\label{eq:deltadef}
   \Delta_{12k}({\tilde e_R^k}) = {|\lambda_{12k}|^2\over 4\sqrt{2} G_F
   M^2_{\tilde e_R^k}},
\end{equation}
with ${\tilde e_R^k}$ being the exchanged slepton, and where 
the $\Delta_{ijk}'({\tilde f})$ are defined similarly by replacing 
$\lambda_{ijk} \to \lambda_{ijk}'$. One obtains tree level contributions to 
$\qwp$ such as those shown in Fig.~\ref{rpvsusy}. Similar corrections affect
other EW observables, such as $\qwe$, $\qwcs$, and $G_V$. 
Specifically~\cite{Ramsey-Musolf:2000qn},
\begin{eqnarray}
\label{rpvshifts}
   \Delta\qwp/\qwp & \approx & \left( {2\over 1 - 4 \sstw} \right)
   [ - 2 \lambda_x \Delta_{12k}(\tilde e_R^k) + 2\Delta_{11k}'(\tilde d_R^k) -
   \Delta_{1j1}'(\tilde q_L^j)], \\
   \Delta\qwe/\qwe & \approx & - \left( {4\over 1 - 4 \sstw} \right)
   \lambda_x \Delta_{12k}(\tilde e^k_R),
\end{eqnarray}
where $\lambda_x = {\hat s}^2 {\hat c}^2 / (1-2{\hat s}^2) \approx 0.33$. 
In contrast to MSSM loop effects, $\qwp$ and $\qwe$ display complementary 
sensitivities to RPV effects. To illustrate, we consider a multi-parameter fit 
to precision data, allowing $\Delta_{12k}$, $\Delta_{11k}'$, $\Delta_{1j1}'$,
and $\Delta_{21k}'$ to be non-zero. The results imply that the possible shifts 
in $\qwp$ and $\qwe$ have opposite relative signs over nearly all the presently
allowed parameter space. We find that shifts of order 
$\Delta\qwp/\qwp \sim 10\%$ are allowed at the 95\%~CL. Thus, a comparison of 
$\qwp$ and $\qwe$ could help distinguish between versions of SUSY with and 
without RPV.

The effects of $\lambda' \neq 0$ are similar to those generated by scalar LQs.
While RPV SUSY provides a natural context in which to discuss the latter, 
vector LQs arise naturally in various GUT 
models~\cite{Pati:vw,Senjanovic:1982ex}. Assuming 
$SU(3)_C \times SU(2)_L \times U(1)_Y$ invariance one obtains 
the Lagrangian~\cite{Buchmuller:1986zs},
\begin{equation}
\ba{lll}
\label{eq:lq}
{\cal L} &=&        h_2^L \bar{u}                       \ell        R_2^L
          +         h_2^R \bar{q}    i \tau_2            e          R_2^R
          +  \tilde{h}_2  \bar{d}                       \ell \tilde{R}_2^L 
\vspace{10pt} \\
         &+&        g_1^L \bar{q}^c  i \tau_2           \ell        S_1^L
          +         g_1^R \bar{u}^c                      e          S_1^R
          +  \tilde{g}_1  \bar{d}^c                      e   \tilde{S}_1^R
          +         g_3   \bar{q}^c  i \tau_2  \vec\tau \ell        S_3 
\vspace{10pt} \\
         &+&        h_1^L \bar{q}   \gamma^\mu          \ell        U_{1\mu}^L
          +         h_1^R \bar{d}   \gamma^\mu           e          U_{1\mu}^R
          +  \tilde{h}_1  \bar{u}   \gamma^\mu           e   \tilde{U}_{1\mu}^R
          +         h_3   \bar{q}   \gamma^\mu \vec\tau \ell        U_{3\mu}
\vspace{10pt}\\
         &+&        g_2^L \bar{d}^c \gamma^\mu          \ell        V_{2\mu}^L
          +         g_2^R \bar{q}^c \gamma^\mu           e          V_{2\mu}^R
          +  \tilde{g}_2  \bar{u}^c \gamma^\mu          \ell \tilde{V}_{2\mu}^L
          +  h.c., 
\ea
\end{equation}
where $q$ and $l$ and the left-handed quark and lepton doublets and $u$, $d$,
and $e$ are the right-handed singlets.  Since we are interested in 
the implications for $\qwp$, we only consider first generation LQs. The first 
two rows in Eq.~(\ref{eq:lq}) involve scalar LQs while the others involve 
vector types. The LQs in the first and third row have fermion number, 
$F = 3 B + L = 0$, while the others have $F = - 2$. The indices refer to 
their isospin representation.

\begin{table}
\begin{center}
\begin{tabular}{|l|c|r||l|c|r|}
\hline
LQ & consistency & $\Delta\qwp/\qwp$ & LQ & consistency & $\Delta\qwp/\qwp$ \\
\hline\hline
$S_1^L$         & $0.57$ & $  9\%$ & $U_{1\mu}^L$         & $0.26$ & $- 8\%$ \\
$S_1^R$         & $0.01$ & $- 6\%$ & $U_{1\mu}^R$         & $0.56$ & $  6\%$ \\
$\tilde{S}_1^R$ & $0.44$ & $- 6\%$ & $\tilde{U}_{1\mu}^R$ & $0.99$ & $ 25\%$ \\
$S_3$           & $0.76$ & $ 10\%$ & $U_{3\mu}$           & $0.31$ & $- 4\%$ \\
\hline
$R_2^L$         & $0.44$ & $-13\%$ & $V_{2\mu}^L$         & $0.87$ & $  9\%$ \\
$R_2^R$         & $0.89$ & $ 15\%$ & $V_{2\mu}^R$         & $0.11$ & $- 7\%$ \\
$\tilde{R}_2^L$ & $0.13$ & $- 4\%$ & $\tilde{V}_{2\mu}^L$ & $0.56$ & $ 14\%$ \\
\hline
\end{tabular}
\end{center}
\caption{Possible impact of LQ interactions on $\qwp$. The left-hand side shows
scalar and the right-hand side vector LQ species. The columns denoted 
{\em consistency\/} give the fractions of the distribution of operator 
coefficients having the same sign as implied by the LQ model. The final columns
give the fractional shifts in $\qwp$ allowed by the data. In more statistical 
terms, {\em consistency\/} is the result of an hypothesis test, while 
the shifts in $\qwp$ reflect parameter estimations that are irrespective of 
the outcome of the hypothesis test.}
\label{tab}
\end{table}

A recent global analysis of scalar LQ constraints from EW data is given in 
Ref.~\cite{Cheung:2001wx}. Here, we extend this analysis to include vector LQ 
interactions. We also update it by including the new $\qwcs$ in 
Eq.~(\ref{eq:qwcs}), hadronic production cross sections at LEP~2 up to 
207~GeV~\cite{Abbaneo:2002}, and the analysis of nuclear $\beta$ decay given
in Ref.~\cite{Towner:2002rg}. We only consider one LQ species at a time. We fit
the data and determine the consistency (shown in Table~\ref{tab}) of the result
with the sign predicted by a given LQ model. The latter is the probability, 
conditional on the data, that the coefficient has the same sign as implied by 
the model. For example, the data favor the presence of $\tilde{U}_{1\mu}^R$, 
while $S_1^R$ is virtually excluded. Assuming a given LQ model, we then 
determine the 95\%~CL upper limit on $\qwp$. Note that this involves 
a renormalization to the physical parameter space of the model. We observe that
the LQ model most favored by the data is ${\tilde U}_{1\mu}^R$ for which 
shifts in $\qwp$ as large as 25\% are allowed. Since the impact of LQs on 
$\qwe$ is loop suppressed, one would not expect it to deviate significantly 
from the SM prediction. Thus, if one observes a large effect in $\qwp$, $\qwe$ 
would serve as a diagnostic tool to distinguish LQ effects from SUSY.

\section{Conclusions}
\label{conclude}

Precise measurements of relatively low energy EW observables continue to play 
an important part in the search for physics beyond the SM. When taken in 
the proper context, such studies can provide unique clues about the nature of 
EW symmetry breaking, grand unification, {\em etc.}. We have shown that 
the weak charge of the proton constitutes a theoretically clean probe of 
new physics. Presently uncalculable, non-perturbative QCD effects are either 
sufficiently small or can be constrained by the current program of parity 
violation measurements so as to render $\qwp$ free from potentially worrisome 
nucleon structure uncertainties. Within the SM, a 4\% determination of $\qwp$ 
--- as planned at JLab --- would yield a 9~$\sigma$ determination of 
the running of the weak mixing angle. Looking beyond the SM, a measurement at 
this level would provide an effective diagnostic tool for new physics, 
particularly when considered in tandem with complementary precision low energy 
studies, such as the SLAC PV M\o ller scattering experiment, cesium APV, 
$a_\mu$, $\beta$ decay, and others. Should future experimental developments 
make an even more precise $\qwp$ measurement possible, the physics impact would
be correspondingly more powerful. Given its theoretical interpretability, 
pursuing such experimental developments appear to be well worth the effort.

\section*{Acknowledgement:}
We thank K.~Cheung, A.~Derevianko, V.~Flambaum, P.~Herczeg, P.~Langacker,
W.~Marciano, S.~Su, and O.~P.~Sushkov for helpful discussions. This work was 
supported by the U.S.~Department of Energy contracts DE--FG02--00ER41146, 
DE--FG03--02ER41215, DE--FG03--00ER4112, by NSF award PHY00--71856, by CONACYT 
(M\'exico) contract 42026--F, and by DGAPA--UNAM contract PAPIIT IN112902.

\end{document}